\documentclass[pra,aps,superscriptaddress,twocolumn,nofootinbib]{revtex4-1}

\usepackage{mathrsfs}
\usepackage{amsfonts}
\usepackage{amssymb}
\usepackage{amsmath}
\usepackage{graphicx}
\usepackage[usenames,dvipsnames]{color}
\usepackage[colorlinks=true,citecolor=blue,linkcolor=magenta]{hyperref}
\usepackage{lmodern}
\usepackage{amsthm}
\usepackage{dsfont,natbib}
\usepackage{subfigure}
\usepackage{textcomp}
\usepackage{tikz}
\usepackage{tikz-3dplot}
\usetikzlibrary{shapes,arrows}
\usepackage{cancel}
\usepackage{makecell}
\usepackage{soul}
\usepackage{physics}
\usepackage{amsfonts}
\usepackage{setspace}
\usepackage{float}
\usepackage{stackengine}
\usepackage{hyperref}

\AtBeginDocument{%
    \newwrite\bibnotes
    \def\bibnotesext{Citations_Export.bib}
    \immediate\openout\bibnotes=\jobname\bibnotesext
    \immediate\write\bibnotes{@CONTROL{REVTEX41Control}}
    \immediate\write\bibnotes{@CONTROL{%
    apsrev41Control,author="08",editor="1",pages="1",title="1",year="1"}}
     \if@filesw
     \immediate\write\@auxout{\string\citation{apsrev41Control}}%
    \fi
}%

\newcommand{\be}{\begin{equation}}
\newcommand{\ee}{\end{equation}}

\begin{document}

\title{ADAPT-QAOA with a classically inspired initial state}

\author{Vishvesha K. Sridhar}
\affiliation{Division of Physics, Mathematics, and Astronomy, Caltech, Pasadena, CA 91125, U.S.A.}
\author{Yanzhu Chen}
\affiliation{Center for Quantum Information Science \& Engineering, Virginia Tech, Blacksburg, VA 24061, U.S.A.}
\affiliation{Department of Physics, Virginia Tech, Blacksburg, VA 24061, U.S.A.}
\author{Bryan Gard}
\affiliation{Georgia Tech Research Institute, Atlanta, GA 30332, USA}
\author{Edwin Barnes}
\affiliation{Center for Quantum Information Science \& Engineering, Virginia Tech, Blacksburg, VA 24061, U.S.A.}
\affiliation{Department of Physics, Virginia Tech, Blacksburg, VA 24061, U.S.A.}
\author{Sophia E. Economou}
\affiliation{Center for Quantum Information Science \& Engineering, Virginia Tech, Blacksburg, VA 24061, U.S.A.}
\affiliation{Department of Physics, Virginia Tech, Blacksburg, VA 24061, U.S.A.}

\date{\today}							

\begin{abstract}
Quantum computing may provide advantage in solving classical optimization problems. One promising algorithm is the quantum approximate optimization algorithm (QAOA). 
There have been many proposals for improving this algorithm, such as using an initial state informed by classical approximation solutions. 
A variation of QAOA called ADAPT-QAOA constructs the ansatz dynamically and can speed up convergence. However, it faces the challenge of frequently converging to excited states which correspond to local minima in the energy landscape, limiting its performance. 
In this work, we propose to start ADAPT-QAOA with an initial state inspired by a classical approximation algorithm. Through numerical simulations we show that this new algorithm can reach the same accuracy with fewer layers than the standard QAOA and the original ADAPT-QAOA. It also appears to be less prone to the problem of converging to excited states.  
\end{abstract}

\maketitle


\section{Introduction}

Quantum-classical hybrid algorithms have been designed to exploit the limited quantum resources at hand by leveraging classical computation. As an example, variational quantum eigensolvers have shown promise in solving problems governed by quantum mechanics, such as finding the ground state energy of molecules~\cite{Peruzzo2014, McClean2016}. In a similar way, classical optimization problems can be solved by mapping the solution to the ground state of a quantum Hamiltonian~\cite{Farhi2002, Lucas2014, Crosson2014, Farhi, Hadfield2017}. It is hoped that allowing the state to explore the Hilbert space will speed up convergence to the solution.

The prototypical algorithm for classical optimization problems is the quantum approximate optimization algorithm (QAOA), which prepares the solution as a parameterized quantum state~\cite{Farhi, Farhi2016, Zhou2020}. 
One example of the problems QAOA is designed to solve is the MaxCut problem, where the goal is to maximize the sum of (possibly weighted) edges on a cut separating the vertices in a graph. 
This problem is NP-hard, where the classical Goemans-Williamson algorithm can obtain an approximation ratio (the cut value found by the algorithm divided by the true solution) of $0.878$ for unweighted graphs, which is the highest guaranteed approximation ratio~\cite{Goemans1995Improved, Tate}.  
As the number of layers in the ansatz approaches infinity, QAOA can reproduce the Trotterized version of an adiabatic evolution to the ground state, and the approximation ratio approaches $1$. However, the performance is limited at a finite number of layers. With just one layer for example, for 3-regular unweighted graphs, the worst-case approximation ratio is $0.6942$~\cite{Farhi}. 

On near-term quantum processors, the circuit depth is limited by decoherence, which prompts the development of algorithms utilizing shallower circuits. It has been shown that a dynamically constructed ansatz in the variational quantum eigensolver can reach the same accuracy with more compact quantum circuits~\cite{Grimsley, Tang2021, Shkolnikov2021}. Similarly, QAOA can benefit from the adaptive strategy in constructing the ansatz. In the algorithm called ADAPT-QAOA, the fixed mixer operator in the ansatz is replaced with an operator selected adaptively in between rounds of optimization~\cite{Zhu}. One challenge in this approach is that the algorithm may find an excited state given the energy gradient criterion used for operator selection, which does not guarantee that the eigenstate the algorithm converges to is the ground state~\cite{Chen2022How}. 
In the adaptive variational quantum eigensolver, this is less of an issue as the optimized state at any step stays close to the global optimum~\cite{Grimsley2023, Anastasiou2022TETRIS}.

Another strategy to further improve QAOA is to use an initial state inspired by classical optimization algorithms. There exist classical algorithms which first relax the rank constraint in the original problem to different extents and later construct a solution by first solving the relaxed problem~\cite{Goemans1995Improved, Burer2003}. In Refs.~\cite{Tate, Egger2021Warm} the authors proposed mapping the solution to the relaxed problem to a quantum state, which is then used as the initial state of QAOA. This so-called warm start improves the performance of a QAOA ansatz with a small number of layers. 
Unlike the standard QAOA, the warm start eliminates the convergence guarantee since the initial state removes the resemblance to adiabatic evolution. This can be remedied by changing the mixer operator so that the initial state is the ground state of the new mixer operator~\cite{Tate2021, Egger2021Warm}. 

For large bounded-degree unweighted graphs, Cain et al. showed that QAOA with a constant depth is unlikely to improve the approximation ratio of a good (but not the optimal) initial state, presenting a challenge for warm starting QAOA~\cite{Cain2022QAOA}. Their analysis applies to QAOA with the original mixer operator and a good computational state as the initial state. In order to circumvent this issue, at least one of these conditions has to be violated. 

In this work, we apply the warm-start strategy in ADAPT-QAOA. We adopt the technique of Refs.~\cite{Tate, Tate2021} and construct the initial state based on the classical solution to a problem whose rank constraint is partially relaxed~\cite{Burer2003}. This initial state has a lower energy expectation value than the standard initial state $\ket{+}^{\otimes n}$, and the standard ADAPT-QAOA procedure will optimize the variational parameters starting from the optimal state from the last layer. In this fashion the energy expectation value of the optimal state at each layer will stay close to the ground state energy. 
We numerically demonstrate the new algorithm, which we call warm-ADAPT-QAOA, on weighted and unweighted regular graphs, and show that it outperforms the standard QAOA and the original ADAPT-QAOA in terms of number of ansatz layers and robustness of finding the ground state. 
We also observe that the approximation ratio of the initial state can be significantly improved by the ansatz.  

The manuscript is organized as follows. In Sec.~\ref{sec:alg} we review both ADAPT-QAOA and the warm-start approach in QAOA, followed by a summary of warm-ADAPT-QAOA. We then demonstrate its performance with numerical simulations for weighted and unweighted graphs in Sec.~\ref{sec:sim}. In Sec.~\ref{sec:first}, we show that the warm-start approach offers improvement for ADAPT-QAOA through analysis of the first step of the two algorithms. We summarize the findings and discuss some open questions in Sec.~\ref{sec:conclusion}.


\section{Description of the algorithm}
\label{sec:alg}

\subsection{Review of ADAPT-QAOA}

Many classical optimization problems can be mapped to finding the ground state of an Ising Hamiltonian~\cite{Farhi}. In this work we focus on the MaxCut problem, where for a $n$-vertex graph the objective function to be maximized is
\be
    F = \frac{1}{2} \sum_{\langle jk \rangle} w_{\langle jk \rangle}({1 - x_jx_k}),
    \label{eq:maxcut}
\ee
where $w_{\langle jk \rangle}$ is the weight of edge $\langle jk \rangle$, and $x_j\in\{\pm1\}$ is the binary variable on the $j$-th vertex. The corresponding Hamiltonian whose energy is to be minimized is 
\be
    C = -\frac{1}{2}\sum_{\langle jk \rangle} w_{\langle jk \rangle}(I - Z_jZ_k),
    \label{eq:cost_ham}
\ee
where $Z_j$ is the Pauli $Z$ operator for the $j$-th qubit. The standard QAOA approximates the ground state with a parameterized ansatz, which is similar to the Trotterized version of the adiabatic evolution~\cite{Farhi},  
\be
    \ket{\psi} = \prod_{i=1}^{p}e^{-i\beta_i M}e^{-i\gamma_i C}\ket{+}^{\otimes n},
\ee
where $p$ is the number of layers and
\be
    M = \sum_{i=1}^{n} X_i.
\ee
This operator is known as the mixer.

Retaining the alternating structure, ADAPT-QAOA replaces the fixed mixer with an operator selected for each layer. It calculates the energy gradient with respect to the new variational parameter for each candidate operator and chooses the one with the largest gradient magnitude~\cite{Zhu}. For a Pauli operator $A$ as the candidate mixer, the gradient is given by
\begin{align}
    &\bra{\psi} e^{i\gamma_0C} i[C, A] e^{-i\gamma_0C} \ket{\Psi} \nonumber \\
    &= 2\mathfrak{Re} \left( \bra{\psi} e^{i\gamma_0C} iC A e^{-i\gamma_0C} \ket{\Psi} \right),
    \label{eq:grad}
\end{align}
where $\ket{\Psi}$ is the current optimal state, $A$ is taken from a pre-selected operator pool and $\gamma_0$ is a small finite value~\cite{Zhu}. All the variational parameters in the new ansatz are subsequently optimized, producing a new optimal state. It is worth mentioning that the optimization at the $k$-th layer starts by initializing the new parameters $\beta_k$ at $0$ and $\gamma_k$ at $\gamma_0$. By adaptively choosing the mixer at each layer from a pool containing two-qubit Pauli operators, ADAPT-QAOA is able to converge to the ground state of the cost function with lower circuit depths and fewer CNOT gates~\cite{Zhu}.

\subsection{Review of warm start}
 
After collecting the set of $n$ binary variables into one $n$-bit variable $x=\{x_j\}\in\{\pm1\}^n$, the MaxCut objective function in Eq.~(\ref{eq:maxcut}) is written as
\begin{align}
    F = \frac{1}{2}W + \frac{1}{4}\text{Tr}(-A^{\rm T} xx^{\rm T}),
\end{align}
where $W = \sum_{\langle jk \rangle}w_{\langle jk \rangle}$ and $A$ is the adjacency matrix of the graph. The $n\times n$ matrix $Y=xx^{\rm T}$ is positive-semidefinite, rank-1, with diagonal entries equal to $1$~\cite{Tate}. Relaxing the rank constraint on $Y$ converts the problem to a semidefinite program~\cite{Goemans1995Improved}. Since $Y$ is positive-semidefinite, one can perform a Cholesky decomposition $Y=X^{\rm T}X$, where each column of $X$ can be identified as a $n$-dimensional vector $\vec{v}_j$ associated with the $j$-th vertex. Each $\vec{v}_j$ has norm $1$ since diagonal entries of $Y$ are $1$. Restoring the rank constraint corresponds to setting the first component of each $\vec{v}_j$ (i.e. $X_{1j}$) to $\pm1$ while leaving all other components $0$. Burer and Monteiro rewrote the relaxed problem as
\begin{align}
    \text{maximize Tr}(-A^{\rm T}X^{\rm T}X), \nonumber \\
    \text{subject to } \abs{\vec{v}_j} = 1 \, \forall j, \nonumber \\
    \vec{v}_j \in \mathbb{R}^n \, \forall j,
\end{align}
where $\vec{v}_j$ is the $j$-th column of $X$ \cite{Burer2003}. They further proposed modifying the constraint $\vec{v}_j \in \mathbb{R}^n$ to $\vec{v}_j \in \mathbb{R}^k$ for some $k<n$, known as a rank-$k$ formulation~\cite{Burer2003}. The original MaxCut problem corresponds to $k=1$ while the $k=n$ relaxation is a semidefinite program. The solution of a relaxed problem $X$ gives a series of $k$-dimensional unit vectors. From this, one can use a hyperplane to produce a cut configuration~\cite{Tate, Burer2003, Goemans1995Improved}. 

Alternatively, the rank-2 or rank-3 Burer-Monteiro solution provides a heuristic initial state for QAOA, known as the warm start, which can improve the performance for the first few layers~\cite{Tate, Tate2021}. Here each of the 2-dimensional or 3-dimensional vectors is interpreted as a quantum state on the Bloch sphere. Since a simultaneous rotation on all the vectors has no influence on the objective in the Burer-Monteiro formulation, such a rotation is performed so that one of the $n$ vectors lands at the north pole of the Bloch sphere. From the $n$ quantum states produced this way, the one with the lowest energy is chosen as the initial state.  

\subsection{Warm-ADAPT-QAOA}

\begin{figure*}
    \centering
    \includegraphics[width=0.75\textwidth]{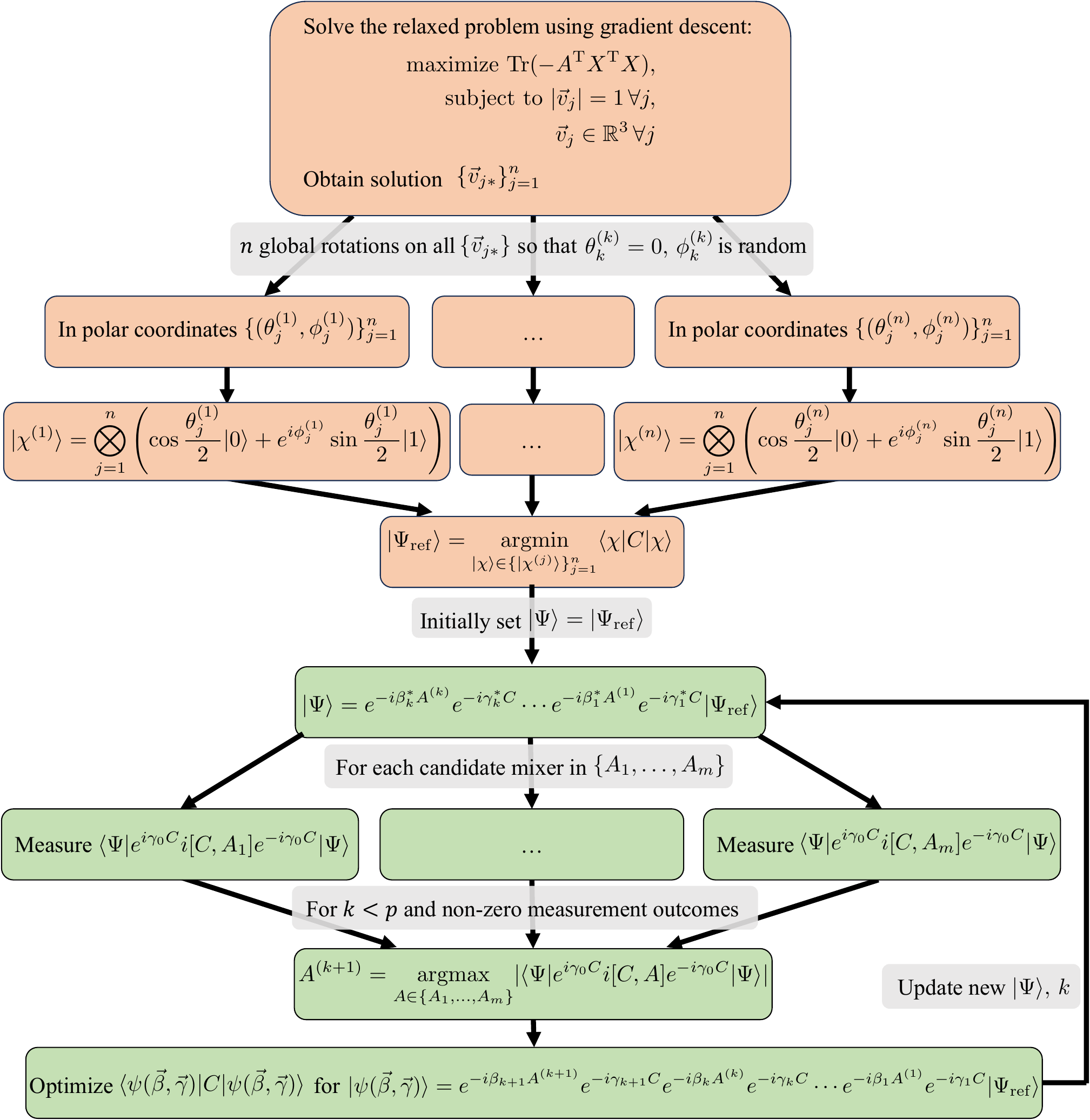}
    \caption{Diagram of warm-ADAPT-QAOA. Orange and green highlight the procedures for finding a warm start and for ADAPT-QAOA, respectively.}
    \label{fig:diag}
\end{figure*}

Following Ref.~\cite{Tate}, we adopt the rank-3 Burer-Monteiro solution and map the simultaneously rotated vectors to a quantum state that will serve as the initial state for ADAPT-QAOA. For an $n$ vertex graph with edge weights $\{w_{\langle jk \rangle}\}$, $n$ random vectors $\{\vec{v}_j\}$ are initialized on a 2-sphere. We then minimize the objective function
\be
    \tilde{F} = \sum_{\langle jk \rangle} w_{\langle jk \rangle} \vec{v}_j \cdot \vec{v}_k,
\ee
by performing stochastic gradient descent on the vectors. All the vectors in the solution are rotated simultaneously according to the criterion mentioned above.
For ADAPT-QAOA, we take the following mixer operator pool $\{\sum_{i=1}^{n} X_i\} \cup \{X_i, Y_i\}_{i = 1, ... n} \cup \{X_j Y_k, X_jZ_k, Y_jZ_k, X_jX_k, Y_jY_k, Z_jZ_k\}_{j,k = 1,...n, j\neq k}$. In addition to the cost Hamiltonian, the two-qubit operators can provide extra entangling operations if selected. 
In Ref.~\cite{Tate2021}, the resemblance to adiabatic evolution is restored by tailoring the mixer operator to the initial state. In a similar way, one can add this adjusted mixer operator to the pool as a candidate, restoring the possibility of realizing adiabatic evolution in the infinite depth limit.  
The adjusted operator takes the form of a sum of single-qubit operators, for which the warm start initial state is the ground state. We denote this version of the algorithm as am-warm-ADAPT-QAOA, where the adjusted mixer is included in the ADAPT operator pool.  
In Fig.~\ref{fig:diag} we summarize the procedure of warm-ADAPT-QAOA.


\section{Numerical simulation}
\label{sec:sim}

Here, we study the performance of warm-ADAPT-QAOA and am-warm-ADAPT-QAOA. The Nelder-Mead optimization method is used to classically optimize the variational parameters $\vec{\gamma}$ and $\vec{\beta}$. We initialize $\gamma$ at $\gamma_0=0.01$ and $\beta$ at $0$, due to $\gamma=0$ being a saddle point of the cost function~\cite{Zhu}. The algorithm is tested on $n$-qubit weighted and unweighted random regular graph instances of degree $D$. 
The weights are drawn from the uniform distribution between 0 and 1. The true ground state energy is calculated exactly and compared to the energy output by the algorithms. The energy error is normalized to the true ground state energy. 
We compare the performance to that of the standard QAOA, the standard QAOA with warm start and a standard mixer, the standard QAOA with warm start along with the adjusted mixer (denoted as am-QAOA warm start), and ADAPT-QAOA. Since it is not our focus to find the best optimizer, we ran each algorithm only once for a given graph, instead of finding the best parameters from multiple optimizers and parameter initialization schemes. We note that this does not guarantee the global energy minimum for a given ansatz is found. 

\subsection{Evidence for Improvement}

\begin{figure*}
\includegraphics[width=\textwidth, height=15cm]{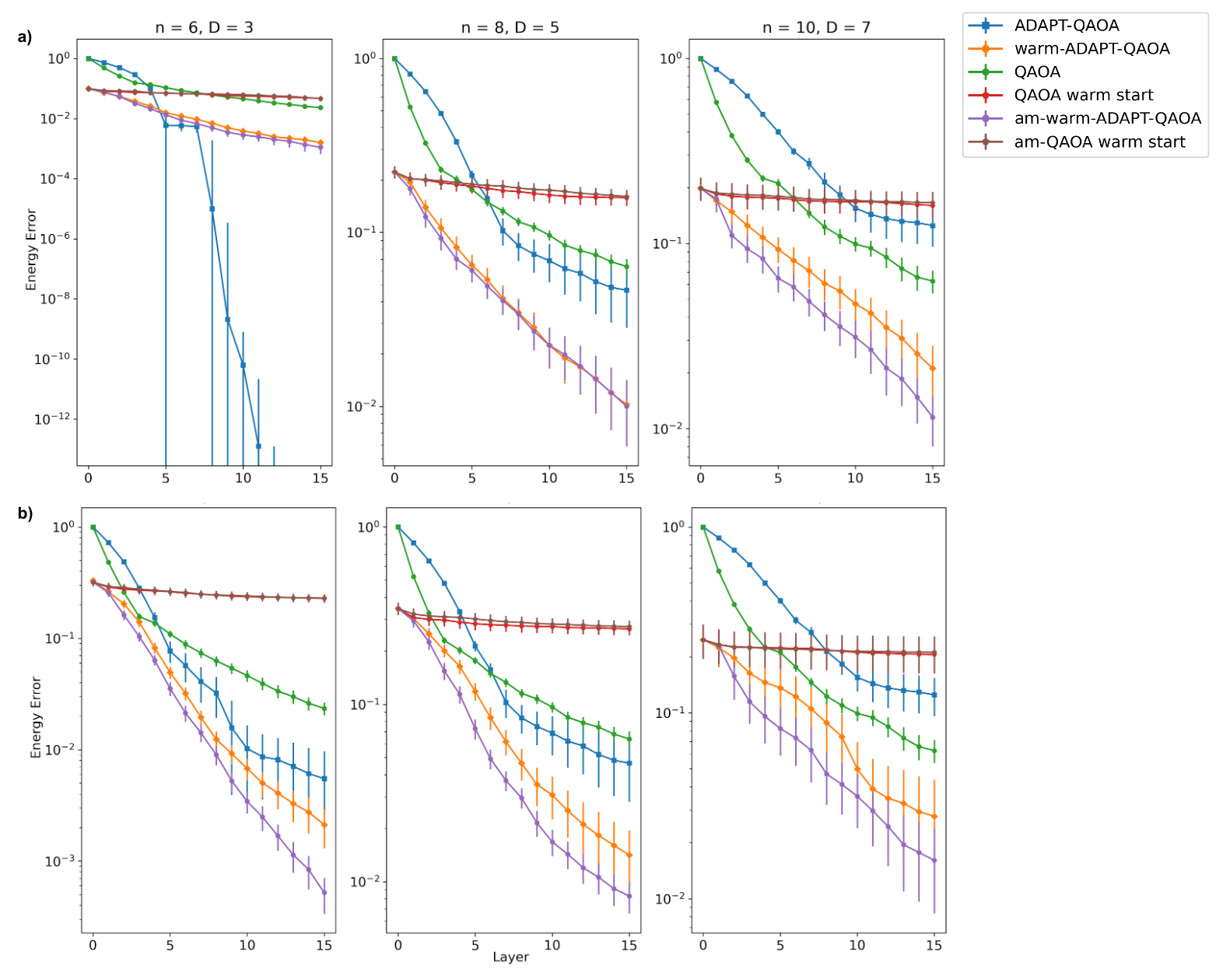}
\caption{Energy error as a function of the number of layers for ADAPT-QAOA, warm-ADAPT-QAOA, the standard QAOA, the standard QAOA with warm start, am-warm-ADAPT-QAOA, and am-QAOA warm start averaged over weighted (a) and unweighted (b) regular graph instances. 40 instances are generated for $n = 6$, 20 instances for $n = 8$, and 10 instances for $n = 10$, where $n$ is the number of vertices and $D$ is the degree of the regular graph. }
\label{fig:performance}
\end{figure*}

In Fig.~\ref{fig:performance} we show the energy error for each layer of the ansatz for $n = 6, D = 3$, $n = 8, D = 5$, and $n = 10, D = 7$ graphs for four different algorithms. We observe that warm-ADAPT-QAOA performs better than the other three algorithms for $n \geq 8$, but not for $n = 6$, where ADAPT-QAOA rapidly converges to the ground state.  
Both warm start algorithms are closer to the true ground state at $p = 1$, but the standard QAOA with warm start and the standard mixer does not significantly improve the energy error from there, corroborating the results of Ref.~\cite{Cain2022QAOA}. On the other hand, warm-ADAPT-QAOA continues to reduce the energy at roughly the same rate as ADAPT-QAOA and the standard QAOA at early layers. For larger graphs, warm-ADAPT-QAOA converges even faster than ADAPT-QAOA at later layers. 
This suggests that warm-ADAPT-QAOA is less prone to the difficulty in further optimization encountered by the standard QAOA with warm start and the standard mixer. 
We also find that compared to warm-ADAPT-QAOA, am-warm-ADAPT-QAOA performs either similarly or slightly better.

\begin{figure}
\includegraphics[width=\columnwidth, height=9cm]{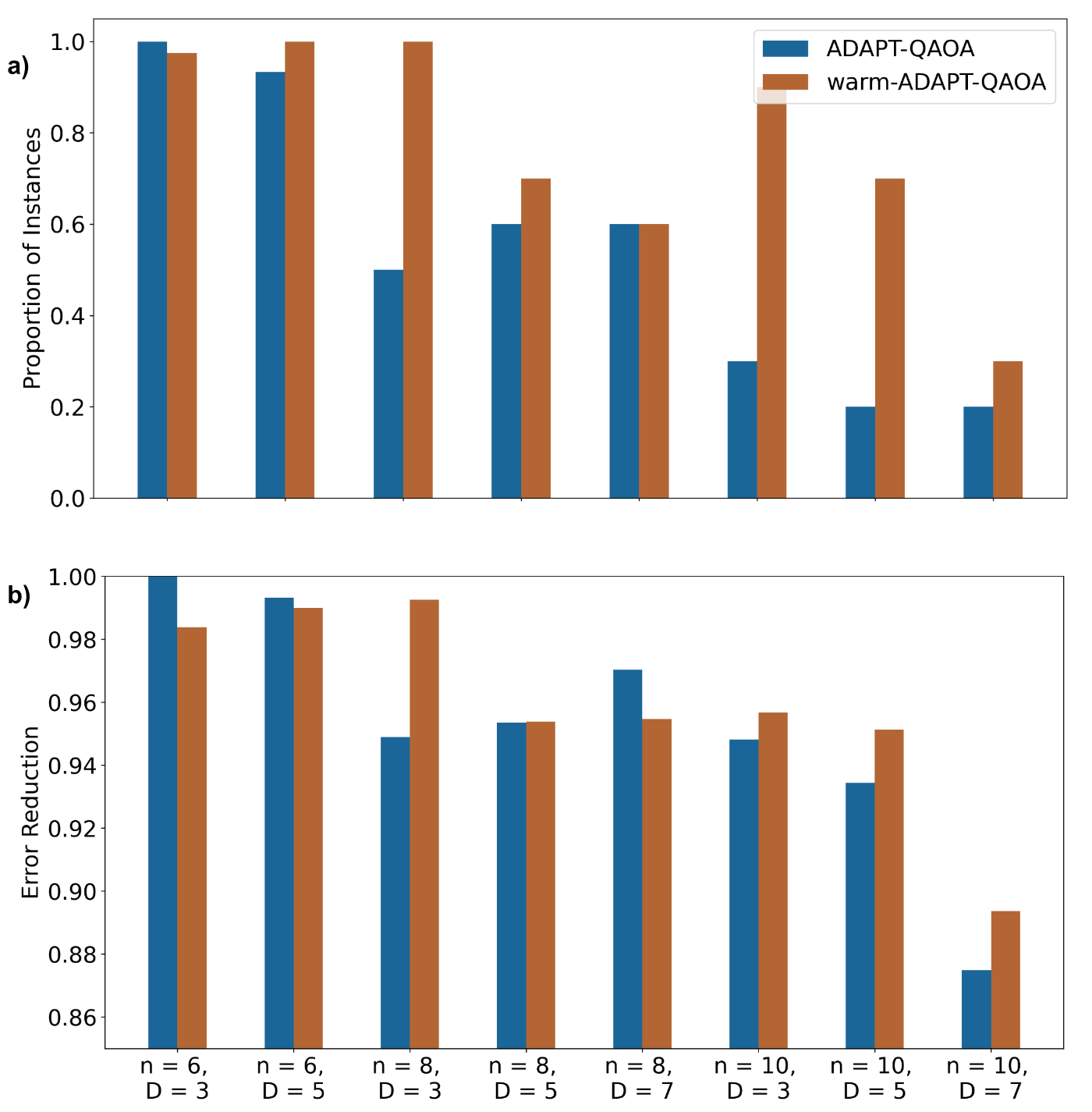}
\caption{a) Proportion of graph instances that reach an energy error of 0.01 within 15 layers. b) Average energy reduction from $p = 0$ to $p = 15$. 40 instances are generated for $n = 6$ graphs, 20 instances for $n = 8$ graphs, 10 instances for $n = 10$ graphs. All graphs are weighted.}
\label{fig:fraction}
\end{figure}

In Fig.~\ref{fig:fraction}a, we present the proportion of graph instances that reach an energy error of $1\%$ within 15 layers. This again shows that warm-ADAPT-QAOA quickly converges to the solution more often for larger graphs, while ADAPT-QAOA slightly outperforms the warm start for smaller graphs.
We further analyze this claim by calculating how much the energy is lowered from $p = 0$ (i.e. the reference state) to $p=15$ for both algorithms, defined as $1-\frac{\langle C\rangle_{p=15}-C_{\rm min}}{\langle C\rangle_{p=0}-C_{\rm min}}$ where $C_{\min}$ is the exact minimum of the cost function. This is plotted in Fig.~\ref{fig:fraction}b. Similarly, for larger graphs, warm-ADAPT-QAOA reaches a much lower energy than ADAPT-QAOA over the course of 15 layers. This indicates that using a warm start is advantageous for larger and more complicated graphs, and may be so in the regime where classical algorithms become inefficient and quantum algorithms can provide speedup.

\begin{figure}
\includegraphics[width=\columnwidth, height=4.7cm]{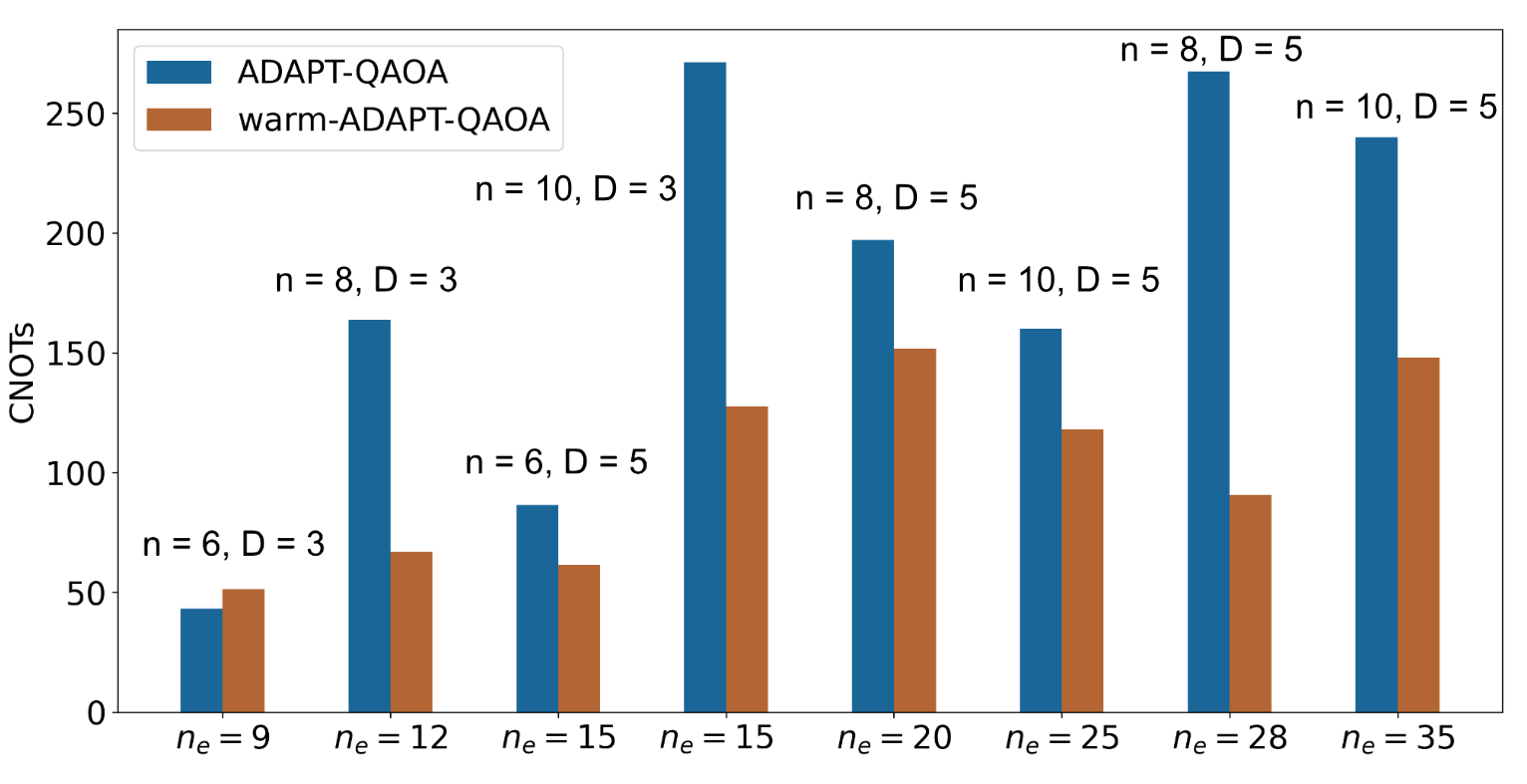}
\caption{Average number of CNOT gates required to reach an energy error of 0.01 for multiple weighted graph instances. The label $n_e$ is the number of edges in the graph. 40 instances are generated for $n=6$ graphs, 20 instances for $n = 8$ graphs, and 10 instances for $n = 10$ graphs.}
\label{fig:resource}
\end{figure}

An important measure of resource cost for near-term quantum algorithms is the number of entangling gates required to obtain a certain accuracy, as these gates are generally more noisy than single-qubit gates. We estimate the number of CNOT gates in the circuit by writing each entangling operation in the ansatz as a combination of two CNOT gates and single qubit gates. In Fig.~\ref{fig:resource} we show the average number of CNOT gates required by each algorithm to reach an energy error of 0.01. It should be noted that this is an upper bound, as further transpilation may reduce the number of CNOT gates. 
For $n\geq8$, warm-ADAPT-QAOA uses significantly fewer CNOT gates than ADAPT-QAOA while for $n = 6$, it uses slightly more. Most of this improvement is due to the warm start algorithm reaching the energy error threshold within fewer layers. Additionally, we note that this average only includes the instances that reach the threshold within 15 layers, when most of the ADAPT-QAOA instances do not achieve this. The plot is a vast underestimate for ADAPT-QAOA, as most instances would require far more layers to achieve the threshold.

\subsection{Dependence on the quality of the warm start}

\begin{figure}
\includegraphics[width=\columnwidth, height=7cm]{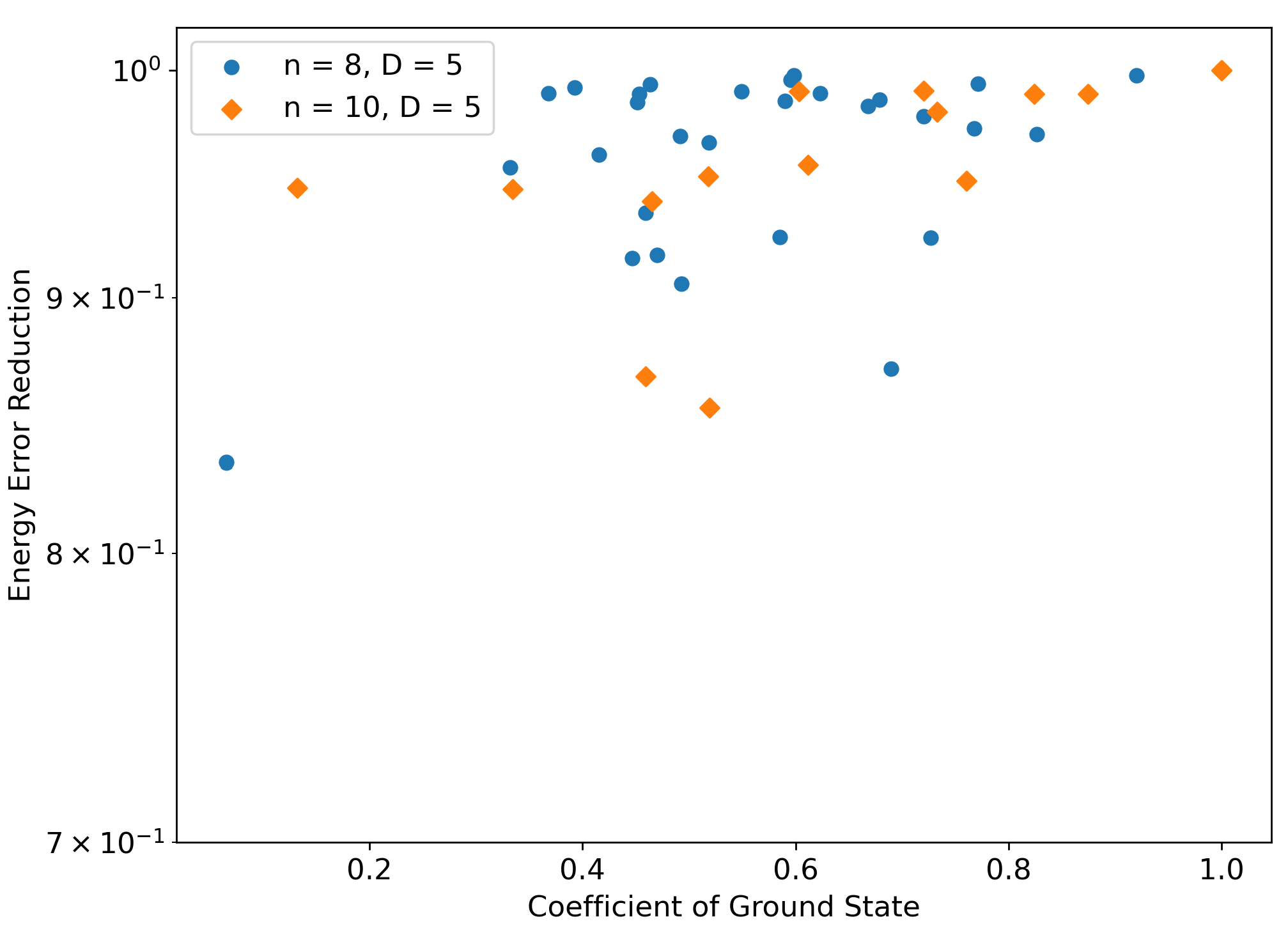}
\caption{Energy error reduction as a function of $\abs{\bra{G}\ket{\psi}}$ for 30 instances of $n=8, D = 5$ weighted graphs and 15 instances of $n=10, D=5$ weighted graphs. }
\label{fig:gs_coeff}
\end{figure}

The warm-start approach can help ADAPT-QAOA circumvent the issue of converging to an excited state by initializing close to the true ground state. However, since the rank-3 Burer-Monteiro relaxation is not a semidefinite program, it is not guaranteed that the solution found by gradient descent is the global minimum of the relaxed objective function, or that the corresponding quantum state is close to the ground state. We therefore analyze how the quality of the warm start affects the performance of warm-ADAPT-QAOA. We quantify the quality of the initial state $\ket{\psi}$ by its overlap $\abs{\bra{G}\ket{\psi}}$ with the true ground state $\ket{G}$. 

For thirty $n = 8$, $D = 5$ weighted graph instances, we find that $63.3\%$ have $\ket{G}$ as the basis state with the largest component in $\ket{\psi}$, and 80\% of fifteen $n = 10, D = 5$ instances also have this property. This indicates that the initial state is usually of high quality.

From Fig.~\ref{fig:gs_coeff} we can see a very weak correlation between the quality of the initial state and the energy error reduction from p = 0 (i.e. the reference state) to p = 15. Although a better initial state seems more likely to give better performance, the algorithm can still significantly improve a relatively poor initial state.


\section{Analysis of the first step}
\label{sec:first}

\subsection{Parameter Landscape}

\begin{figure*}
\includegraphics[width=\textwidth, height=9cm]{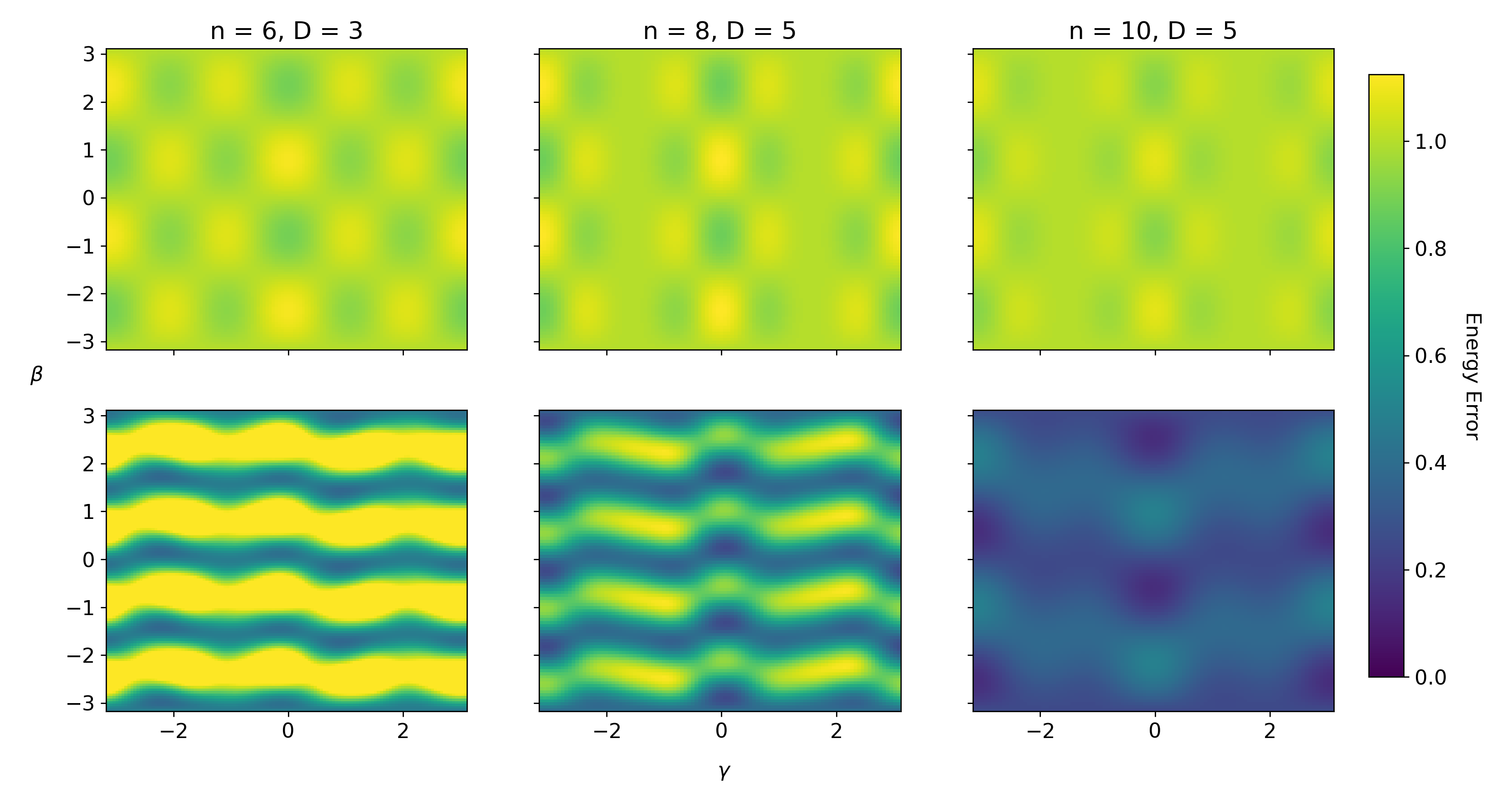}
\caption{Parameter landscape for different unweighted ADAPT landscapes are shown in top row, warm-ADAPT landscapes shown in bottom row.}
\label{fig:landscape}
\end{figure*}

To further analyze the difference between ADAPT-QAOA and warm-ADAPT-QAOA, we analyze the complete space of possible parameters at $p = 1$. We plot the energy error at every point in a grid of $\gamma$ and $\beta$ values for a random instance of an unweighted 5-regular graph, as in Fig.~\ref{fig:landscape}. We perform this analysis for $n = 6, 8, 10$. This indicates how difficult classical optimization is at $p = 1$.

We observe large differences between the parameter landscapes of warm-ADAPT-QAOA and the original ADAPT-QAOA. The standard-start landscape consists of peaks and valleys, while warm-ADAPT-QAOA creates a landscape of multiple ridges. This makes warm-ADAPT-QAOA less sensitive to the initial choice of parameters. In particular, since $\beta$ and $\gamma$ are initialized at $0$ and a small value respectively, classical optimization starts from a trough. 

As the graph size increases with the same connectivity, the ADAPT-QAOA landscape generally gets flatter with higher valleys and lower peaks, although for  $n =6, D=3$, this trend does not necessarily hold. For $n=6, 8, 10$, the minimum energy errors are $\frac{8}{9}, \frac{7}{8}, \frac{12}{13}$, respectively, and the average energy error is constant at $1$. The warm start landscape also exhibits lower maxima as graph size increases. The minimum energy errors are $10^{-6}$, $0.249$, and $0.145$ for $n=6, 8, 10$. For the $n = 6$ graph, the warm start itself is very close to the solution. The average energy errors for the warm start are $0.5$, $0.673$, and $0.316$, respectively. 
This indicates that warm-ADAPT-QAOA is less affected by an increase in the graph size than ADAPT-QAOA and may help explain why warm-ADAPT-QAOA performs better for larger graphs.  
Overall, the first layer parameter landscape of the warm-start algorithm is more favorable than that of standard ADAPT-QAOA.

\subsection{ADAPT-QAOA first step}

Although ADAPT-QAOA can offer improvement over the standard QAOA for some number of layers, its first layer performance is typically worse. 
With the initial state $\ket{+}^{\otimes n}$, the leading term in the gradient given by Eq.~(\ref{eq:grad}) is proportional to $\gamma_0$ for any operator in the chosen pool except those of the form $Y_jZ_k$. For $Y_jZ_k$, the leading term in the gradient magnitude is $w_{\langle jk \rangle}$. With a sufficiently small $\gamma_0\ll1$ and the edge weights drawn between $0$ and $1$, the mixer operator selected at the first layer is most likely to be $Y_jZ_k$ with the largest $w_{\langle jk \rangle}$. For an unweighted graph whose edge weights are all set to $1$, whether a $Y_jZ_k$ with a connected edge $\langle jk \rangle$ or another operator gets selected depends on the size of the graph as well as the value of $\gamma_0$. 

We can thus analyze the behavior of ADAPT-QAOA with $p=1$ in some special cases. We take the unweighted regular graphs as simple examples. 
After numerically optimizing the parameters, we obtain the maximum cut value $\frac{nD + 2}{4}$ at $\gamma$ near 0 and $\beta$ near $\frac{\pi}{4}$.   
For rings where $D = 2$, the magnitude of the largest cut ADAPT-QAOA can find at $p = 1$ is $\frac{n+1}{2}$. It was shown that standard QAOA can find a cut of size $\frac{3n}{4}$~\cite{Farhi}. This is larger than the ADAPT-QAOA cut for all $n > 1$. 
For 3-regular graphs, the minimum $p = 1$ approximation ratio of the standard QAOA is $0.6924$ and the largest possible MaxCut value for any graph is $\frac{3n}{2}$~\cite{Farhi}. The approximation ratio of $p = 1$ ADAPT-QAOA for such graphs is $\frac{3n + 2}{6n}$, lower than that of the standard QAOA for all $n > 1$.

\subsection{warm-ADAPT first step}

\begin{figure}
\includegraphics[width=\columnwidth, height=6cm]{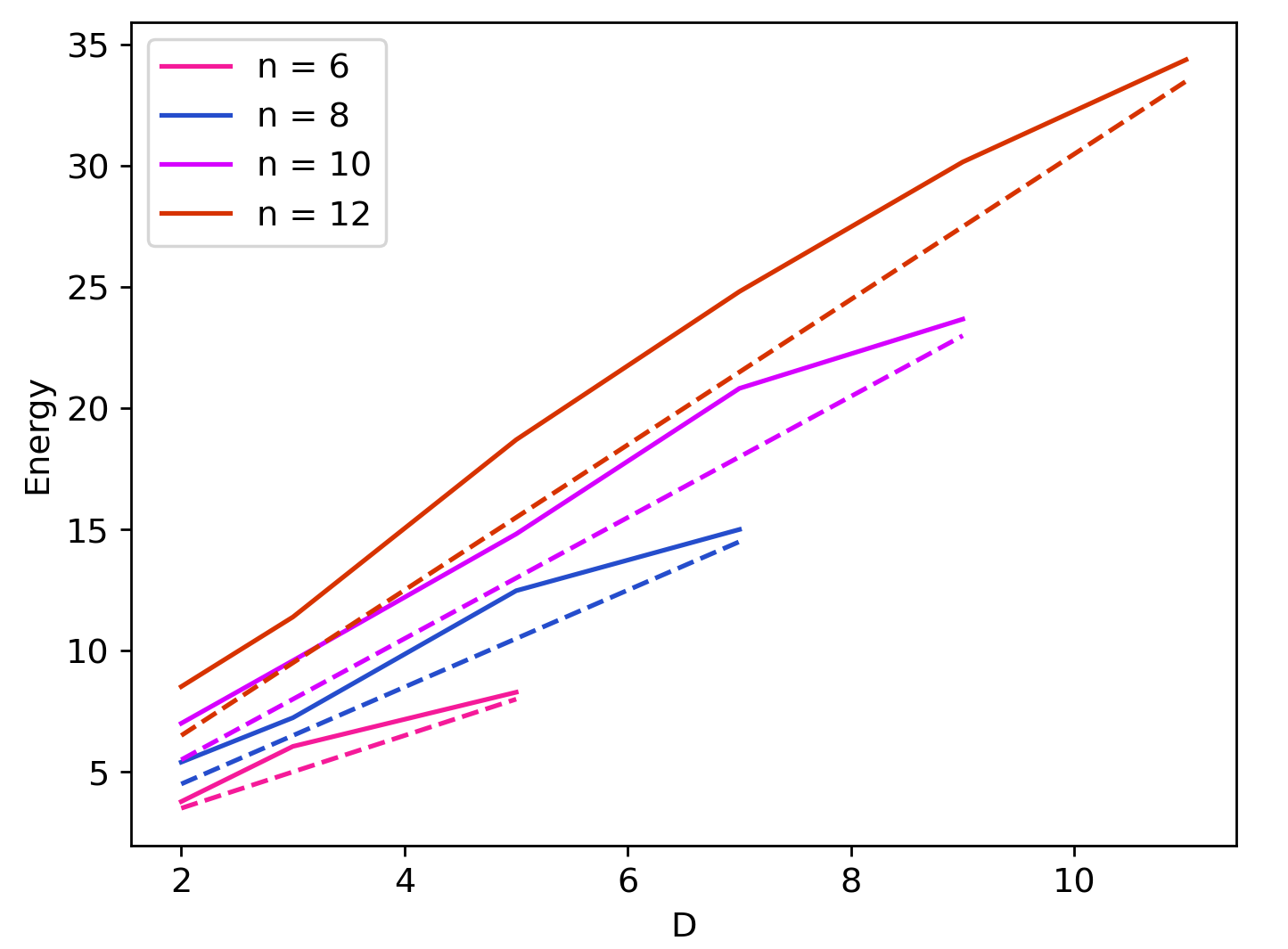}
\caption{Minimum cut values, returned by warm-ADAPT-QAOA out of 75 unweighted regular graph instances, are shown in solid lines. Dashed lines indicate the maximum cut values that ADAPT-QAOA can return. }
\label{fig:first}
\end{figure}

\begin{figure*}
\includegraphics[width=\linewidth, height=9.5cm]{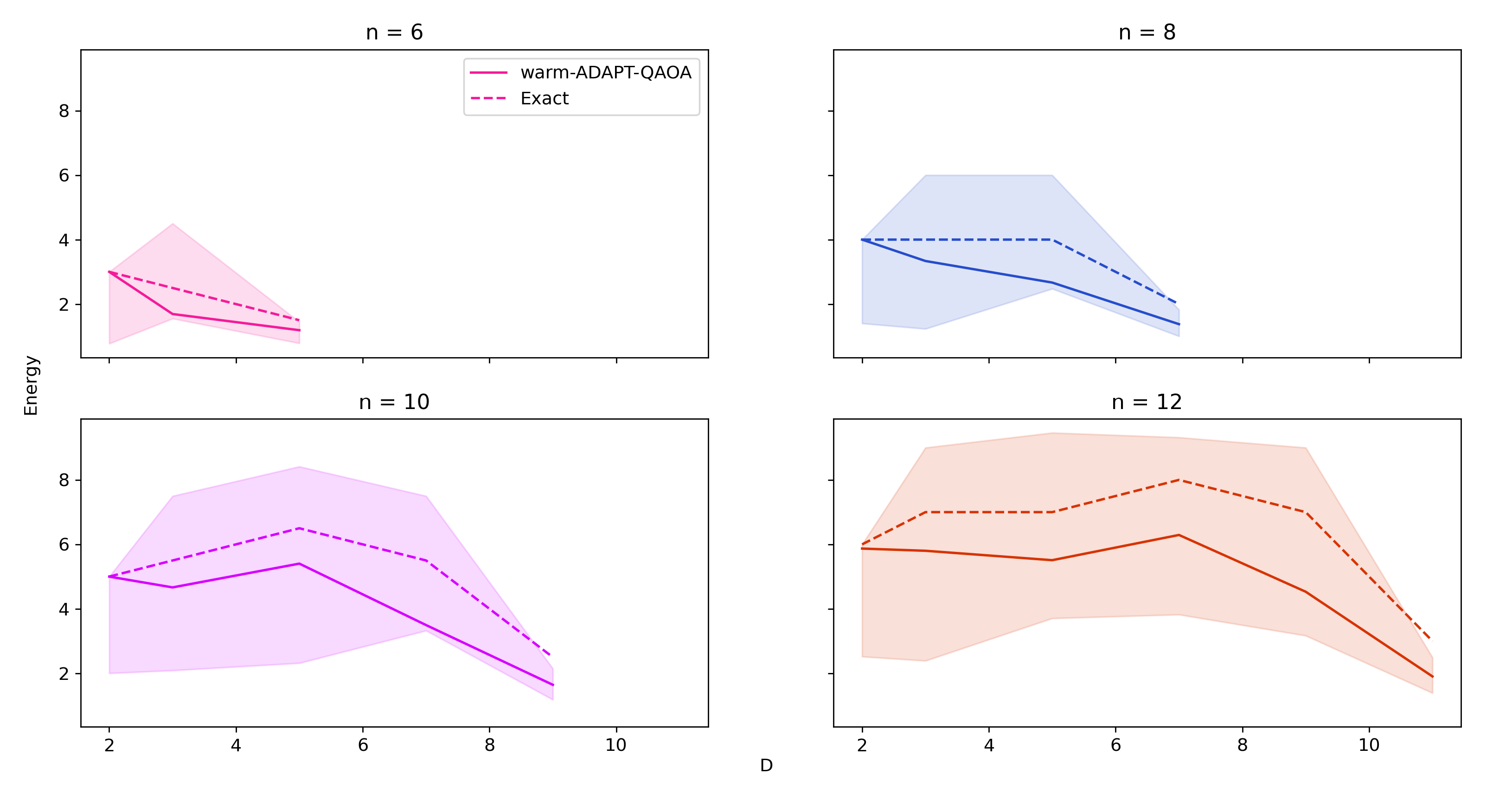}
\caption{Median energy expectation value returned by warm-ADAPT-QAOA out of 75 unweighted regular graph instances of various sizes and connectivities shown in solid line, median exact maximum energy for the same graphs shown in dotted line. The shaded area indicates values between the maximum and minimum energy expectations values returned by warm-ADAPT-QAOA from the 75 instances. The constant term in Eq.~\ref{eq:cost_ham} is dropped here.}
\label{fig:first_max_med_min}
\end{figure*}

With different possible initial states, warm-ADAPT-QAOA does not always choose the same operator in the first layer, so it is more difficult to analyze its performance. Here, we numerically examine the first layer of 75 unweighted regular graph instances of various sizes and connectivities. 
In Fig.~\ref{fig:first}, we observe that even over 75 instances, ADAPT-QAOA never returns a higher cut value than warm-ADAPT-QAOA at the first layer. This allows us to reasonably conclude that warm-ADAPT-QAOA will always outperform ADAPT-QAOA at the first layer. For lower connectivities, the median energy is much higher than the minimum energy, as shown in Fig.~\ref{fig:first_max_med_min}. This means that for a majority of instances, warm-ADAPT-QAOA will return a cut whose magnitude far exceeds that of the cut returned by ADAPT-QAOA. For $D = 3$, the largest cut value out of the 75 instances is $\frac{3n}{2}$ to within 8 decimal places, which means that warm-ADAPT-QAOA is able to return the largest cut for a 3-regular graph within one layer. 

\section{Conclusion and discussion}
\label{sec:conclusion}

We propose a variation of ADAPT-QAOA that starts with an initial state inspired by a classical approximation algorithm. 
We numerically simulate its performance and show that on average warm-ADAPT-QAOA can reach a better accuracy than ADAPT-QAOA with the same number of layers in the ansatz. Consequently, it requires fewer resources. 
Since the initial state typically has a lower energy than the state $\ket{+}^{\otimes n}$, it may be unsurprising that this new algorithm significantly outperforms ADAPT-QAOA at the first layer. However, in subsequent layers, it can provide a significant reduction in energy, contrary to the challenge observed in the standard QAOA with warm start~\cite{Cain2022QAOA}. This indicates that ADAPT-QAOA, with its problem-tailored ansatz, is more compatible with the warm-start approach. 

We also see that adding the mixer operator adjusted to the warm-start initial state to the operator pool may improve the performance further. In the standard QAOA with warm start, the adjusted mixer plays the role of recovering the resemblance to adiabatic evolution. The connection between ADAPT-QAOA and the shortcut to adiabaticity has been discussed in Ref.~\cite{Zhu}. One can speculate that by including the adjusted mixer in the operator pool, ADAPT-QAOA could approach the shortcut to adiabaticity with the warm-start initial state better. While in this work we primarily focus on the effect of a warm start on ADAPT-QAOA, without extensively studying the role of the adjusted mixer operator, we believe this is an interesting direction for future work, which can shed light on how to choose the initial state and the operator pool in a compatible and optimal way.   

In the simulations, we observe that by starting from a state with a significant overlap with the ground state, the new approach appears to circumvent the issue of converging to excited states. It remains an open question how our new approach scales to larger problem sizes, and whether it provides some robustness against local minima of the energy landscape.


\begin{acknowledgements}
S. E. E. acknowledges support from the US Department of Energy (Award No. DE-SC0019318). E. B. acknowledges support from the US Department of Energy (Award No. DE-SC0019199). 
\end{acknowledgements}

\bibliographystyle{apsrev}	
\bibliography{warm-adapt-qaoa}

\begin{thebibliography}{22}
\expandafter\ifx\csname natexlab\endcsname\relax\def\natexlab#1{#1}\fi
\expandafter\ifx\csname bibnamefont\endcsname\relax
  \def\bibnamefont#1{#1}\fi
\expandafter\ifx\csname bibfnamefont\endcsname\relax
  \def\bibfnamefont#1{#1}\fi
\expandafter\ifx\csname citenamefont\endcsname\relax
  \def\citenamefont#1{#1}\fi
\expandafter\ifx\csname url\endcsname\relax
  \def\url#1{\texttt{#1}}\fi
\expandafter\ifx\csname urlprefix\endcsname\relax\def\urlprefix{URL }\fi
\providecommand{\bibinfo}[2]{#2}
\providecommand{\eprint}[2][]{\url{#2}}

\bibitem[{\citenamefont{Peruzzo et~al.}(2014)\citenamefont{Peruzzo, McClean,
  Shadbolt, Yung, Zhou, Love, Aspuru-Guzik, and O'Brien}}]{Peruzzo2014}
\bibinfo{author}{\bibfnamefont{A.}~\bibnamefont{Peruzzo}},
  \bibinfo{author}{\bibfnamefont{J.}~\bibnamefont{McClean}},
  \bibinfo{author}{\bibfnamefont{P.}~\bibnamefont{Shadbolt}},
  \bibinfo{author}{\bibfnamefont{M.-H.} \bibnamefont{Yung}},
  \bibinfo{author}{\bibfnamefont{X.-Q.} \bibnamefont{Zhou}},
  \bibinfo{author}{\bibfnamefont{P.~J.} \bibnamefont{Love}},
  \bibinfo{author}{\bibfnamefont{A.}~\bibnamefont{Aspuru-Guzik}},
  \bibnamefont{and} \bibinfo{author}{\bibfnamefont{J.~L.}
  \bibnamefont{O'Brien}}, \bibinfo{journal}{Nat. Phys.}
  \textbf{\bibinfo{volume}{5}}, \bibinfo{pages}{4213} (\bibinfo{year}{2014}).

\bibitem[{\citenamefont{McClean et~al.}(2016)\citenamefont{McClean, Romero,
  Babbush, and Aspuru-Guzik}}]{McClean2016}
\bibinfo{author}{\bibfnamefont{J.~R.} \bibnamefont{McClean}},
  \bibinfo{author}{\bibfnamefont{J.}~\bibnamefont{Romero}},
  \bibinfo{author}{\bibfnamefont{R.}~\bibnamefont{Babbush}}, \bibnamefont{and}
  \bibinfo{author}{\bibfnamefont{A.}~\bibnamefont{Aspuru-Guzik}},
  \bibinfo{journal}{New J. Phys.} \textbf{\bibinfo{volume}{18}},
  \bibinfo{pages}{023023} (\bibinfo{year}{2016}).

\bibitem[{\citenamefont{{Farhi} et~al.}(2002)\citenamefont{{Farhi},
  {Goldstone}, and {Gutmann}}}]{Farhi2002}
\bibinfo{author}{\bibfnamefont{E.}~\bibnamefont{{Farhi}}},
  \bibinfo{author}{\bibfnamefont{J.}~\bibnamefont{{Goldstone}}},
  \bibnamefont{and}
  \bibinfo{author}{\bibfnamefont{S.}~\bibnamefont{{Gutmann}}},
  \bibinfo{journal}{quant-ph/0201031}  (\bibinfo{year}{2002}).

\bibitem[{\citenamefont{Lucas}(2014)}]{Lucas2014}
\bibinfo{author}{\bibfnamefont{A.}~\bibnamefont{Lucas}},
  \bibinfo{journal}{Frontiers in Physics} \textbf{\bibinfo{volume}{2}},
  \bibinfo{pages}{5} (\bibinfo{year}{2014}).

\bibitem[{\citenamefont{{Crosson} et~al.}(2014)\citenamefont{{Crosson},
  {Farhi}, {Yen-Yu Lin}, {Lin}, and {Shor}}}]{Crosson2014}
\bibinfo{author}{\bibfnamefont{E.}~\bibnamefont{{Crosson}}},
  \bibinfo{author}{\bibfnamefont{E.}~\bibnamefont{{Farhi}}},
  \bibinfo{author}{\bibfnamefont{C.}~\bibnamefont{{Yen-Yu Lin}}},
  \bibinfo{author}{\bibfnamefont{H.-H.} \bibnamefont{{Lin}}}, \bibnamefont{and}
  \bibinfo{author}{\bibfnamefont{P.}~\bibnamefont{{Shor}}},
  \bibinfo{journal}{arXiv:1401.7320}  (\bibinfo{year}{2014}).

\bibitem[{\citenamefont{Farhi and Goldstone}(2014)}]{Farhi}
\bibinfo{author}{\bibfnamefont{E.}~\bibnamefont{Farhi}} \bibnamefont{and}
  \bibinfo{author}{\bibfnamefont{J.}~\bibnamefont{Goldstone}},
  \bibinfo{journal}{arXiv:1411.4028}  (\bibinfo{year}{2014}).

\bibitem[{\citenamefont{Hadfield et~al.}(2017)\citenamefont{Hadfield, Wang,
  Rieffel, O'Gorman, Venturelli, and Biswas}}]{Hadfield2017}
\bibinfo{author}{\bibfnamefont{S.}~\bibnamefont{Hadfield}},
  \bibinfo{author}{\bibfnamefont{Z.}~\bibnamefont{Wang}},
  \bibinfo{author}{\bibfnamefont{E.~G.} \bibnamefont{Rieffel}},
  \bibinfo{author}{\bibfnamefont{B.}~\bibnamefont{O'Gorman}},
  \bibinfo{author}{\bibfnamefont{D.}~\bibnamefont{Venturelli}},
  \bibnamefont{and} \bibinfo{author}{\bibfnamefont{R.}~\bibnamefont{Biswas}},
  in \emph{\bibinfo{booktitle}{Proceedings of the Second International Workshop
  on Post Moores Era Supercomputing}} (\bibinfo{publisher}{Association for
  Computing Machinery}, \bibinfo{address}{New York, NY, USA},
  \bibinfo{year}{2017}), PMES'17, p. \bibinfo{pages}{15–21}, ISBN
  \bibinfo{isbn}{9781450351263}.

\bibitem[{\citenamefont{Farhi and Harrow}(2016)}]{Farhi2016}
\bibinfo{author}{\bibfnamefont{E.}~\bibnamefont{Farhi}} \bibnamefont{and}
  \bibinfo{author}{\bibfnamefont{A.~W.} \bibnamefont{Harrow}},
  \bibinfo{journal}{arXiv:1602.07674}  (\bibinfo{year}{2016}).

\bibitem[{\citenamefont{Zhou et~al.}(2020)\citenamefont{Zhou, Wang, Choi,
  Pichler, and Lukin}}]{Zhou2020}
\bibinfo{author}{\bibfnamefont{L.}~\bibnamefont{Zhou}},
  \bibinfo{author}{\bibfnamefont{S.-T.} \bibnamefont{Wang}},
  \bibinfo{author}{\bibfnamefont{S.}~\bibnamefont{Choi}},
  \bibinfo{author}{\bibfnamefont{H.}~\bibnamefont{Pichler}}, \bibnamefont{and}
  \bibinfo{author}{\bibfnamefont{M.~D.} \bibnamefont{Lukin}},
  \bibinfo{journal}{Phys. Rev. X} \textbf{\bibinfo{volume}{10}},
  \bibinfo{pages}{021067} (\bibinfo{year}{2020}).

\bibitem[{\citenamefont{Goemans and Williamson}(1995)}]{Goemans1995Improved}
\bibinfo{author}{\bibfnamefont{M.~X.} \bibnamefont{Goemans}} \bibnamefont{and}
  \bibinfo{author}{\bibfnamefont{D.~P.} \bibnamefont{Williamson}},
  \bibinfo{journal}{Journal of the ACM} \textbf{\bibinfo{volume}{42}},
  \bibinfo{pages}{1115} (\bibinfo{year}{1995}).

\bibitem[{\citenamefont{Tate et~al.}(2022)\citenamefont{Tate, Farhadi, Herold,
  Mohler, and Gupta}}]{Tate}
\bibinfo{author}{\bibfnamefont{R.}~\bibnamefont{Tate}},
  \bibinfo{author}{\bibfnamefont{M.}~\bibnamefont{Farhadi}},
  \bibinfo{author}{\bibfnamefont{C.}~\bibnamefont{Herold}},
  \bibinfo{author}{\bibfnamefont{G.}~\bibnamefont{Mohler}}, \bibnamefont{and}
  \bibinfo{author}{\bibfnamefont{S.}~\bibnamefont{Gupta}},
  \bibinfo{journal}{ACM Transactions on Quantum Computing}
  (\bibinfo{year}{2022}).

\bibitem[{\citenamefont{Grimsley et~al.}(2019)\citenamefont{Grimsley, Economou,
  Barnes, and Mayhall}}]{Grimsley}
\bibinfo{author}{\bibfnamefont{H.~R.} \bibnamefont{Grimsley}},
  \bibinfo{author}{\bibfnamefont{S.~E.} \bibnamefont{Economou}},
  \bibinfo{author}{\bibfnamefont{E.}~\bibnamefont{Barnes}}, \bibnamefont{and}
  \bibinfo{author}{\bibfnamefont{N.~J.} \bibnamefont{Mayhall}},
  \bibinfo{journal}{Nature Communications} \textbf{\bibinfo{volume}{10}}
  (\bibinfo{year}{2019}).

\bibitem[{\citenamefont{Tang et~al.}(2021)\citenamefont{Tang, Shkolnikov,
  Barron, Grimsley, Mayhall, Barnes, and Economou}}]{Tang2021}
\bibinfo{author}{\bibfnamefont{H.~L.} \bibnamefont{Tang}},
  \bibinfo{author}{\bibfnamefont{V.}~\bibnamefont{Shkolnikov}},
  \bibinfo{author}{\bibfnamefont{G.~S.} \bibnamefont{Barron}},
  \bibinfo{author}{\bibfnamefont{H.~R.} \bibnamefont{Grimsley}},
  \bibinfo{author}{\bibfnamefont{N.~J.} \bibnamefont{Mayhall}},
  \bibinfo{author}{\bibfnamefont{E.}~\bibnamefont{Barnes}}, \bibnamefont{and}
  \bibinfo{author}{\bibfnamefont{S.~E.} \bibnamefont{Economou}},
  \bibinfo{journal}{PRX Quantum} \textbf{\bibinfo{volume}{2}},
  \bibinfo{pages}{020310} (\bibinfo{year}{2021}).

\bibitem[{\citenamefont{Shkolnikov et~al.}(2023)\citenamefont{Shkolnikov,
  Mayhall, Economou, and Barnes}}]{Shkolnikov2021}
\bibinfo{author}{\bibfnamefont{V.~O.} \bibnamefont{Shkolnikov}},
  \bibinfo{author}{\bibfnamefont{N.~J.} \bibnamefont{Mayhall}},
  \bibinfo{author}{\bibfnamefont{S.~E.} \bibnamefont{Economou}},
  \bibnamefont{and} \bibinfo{author}{\bibfnamefont{E.}~\bibnamefont{Barnes}},
  \bibinfo{journal}{{Quantum}} \textbf{\bibinfo{volume}{7}},
  \bibinfo{pages}{1040} (\bibinfo{year}{2023}), ISSN \bibinfo{issn}{2521-327X},
  \urlprefix\url{https://doi.org/10.22331/q-2023-06-12-1040}.

\bibitem[{\citenamefont{Zhu et~al.}(2022)\citenamefont{Zhu, Tang, Barron,
  Calderon-Vargas, Mayhall, Barnes, and Economou}}]{Zhu}
\bibinfo{author}{\bibfnamefont{L.}~\bibnamefont{Zhu}},
  \bibinfo{author}{\bibfnamefont{H.~L.} \bibnamefont{Tang}},
  \bibinfo{author}{\bibfnamefont{G.~S.} \bibnamefont{Barron}},
  \bibinfo{author}{\bibfnamefont{F.~A.} \bibnamefont{Calderon-Vargas}},
  \bibinfo{author}{\bibfnamefont{N.~J.} \bibnamefont{Mayhall}},
  \bibinfo{author}{\bibfnamefont{E.}~\bibnamefont{Barnes}}, \bibnamefont{and}
  \bibinfo{author}{\bibfnamefont{S.~E.} \bibnamefont{Economou}},
  \bibinfo{journal}{Phys. Rev. Res.} \textbf{\bibinfo{volume}{4}},
  \bibinfo{pages}{033029} (\bibinfo{year}{2022}),
  \urlprefix\url{https://link.aps.org/doi/10.1103/PhysRevResearch.4.033029}.

\bibitem[{\citenamefont{Chen et~al.}(2022)\citenamefont{Chen, Zhu, Liu,
  Mayhall, Barnes, and Economou}}]{Chen2022How}
\bibinfo{author}{\bibfnamefont{Y.}~\bibnamefont{Chen}},
  \bibinfo{author}{\bibfnamefont{L.}~\bibnamefont{Zhu}},
  \bibinfo{author}{\bibfnamefont{C.}~\bibnamefont{Liu}},
  \bibinfo{author}{\bibfnamefont{N.}~\bibnamefont{Mayhall}},
  \bibinfo{author}{\bibfnamefont{E.}~\bibnamefont{Barnes}}, \bibnamefont{and}
  \bibinfo{author}{\bibfnamefont{S.}~\bibnamefont{Economou}},
  \bibinfo{journal}{arXiv:2205.12283}  (\bibinfo{year}{2022}).

\bibitem[{\citenamefont{Grimsley et~al.}(2023))\citenamefont{Grimsley, Barron,
  Barnes, Economou, and Mayhall}}]{Grimsley2023}
\bibinfo{author}{\bibfnamefont{H.~R.} \bibnamefont{Grimsley}},
  \bibinfo{author}{\bibfnamefont{G.~S.} \bibnamefont{Barron}},
  \bibinfo{author}{\bibfnamefont{E.}~\bibnamefont{Barnes}},
  \bibinfo{author}{\bibfnamefont{S.~E.} \bibnamefont{Economou}},
  \bibnamefont{and} \bibinfo{author}{\bibfnamefont{N.~J.}
  \bibnamefont{Mayhall}}, \bibinfo{journal}{npj Quantum Inf}
  \textbf{\bibinfo{volume}{9}}, \bibinfo{pages}{19} (\bibinfo{year}{2023)}).

\bibitem[{\citenamefont{Anastasiou et~al.}(2022)\citenamefont{Anastasiou, Chen,
  Mayhall, Barnes, and Economou}}]{Anastasiou2022TETRIS}
\bibinfo{author}{\bibfnamefont{P.~G.} \bibnamefont{Anastasiou}},
  \bibinfo{author}{\bibfnamefont{Y.}~\bibnamefont{Chen}},
  \bibinfo{author}{\bibfnamefont{N.~J.} \bibnamefont{Mayhall}},
  \bibinfo{author}{\bibfnamefont{E.}~\bibnamefont{Barnes}}, \bibnamefont{and}
  \bibinfo{author}{\bibfnamefont{S.~E.} \bibnamefont{Economou}},
  \bibinfo{journal}{arXiv:2209.10562}  (\bibinfo{year}{2022}).

\bibitem[{\citenamefont{Burer and Monteiro}(2003)}]{Burer2003}
\bibinfo{author}{\bibfnamefont{S.}~\bibnamefont{Burer}} \bibnamefont{and}
  \bibinfo{author}{\bibfnamefont{R.}~\bibnamefont{Monteiro}},
  \bibinfo{journal}{Math. Program.} \textbf{\bibinfo{volume}{Ser. B}},
  \bibinfo{pages}{329–357} (\bibinfo{year}{2003}).

\bibitem[{\citenamefont{Egger et~al.}(2021)\citenamefont{Egger, Mare{\v c}ek,
  and Woerner}}]{Egger2021Warm}
\bibinfo{author}{\bibfnamefont{D.~J.} \bibnamefont{Egger}},
  \bibinfo{author}{\bibfnamefont{J.}~\bibnamefont{Mare{\v c}ek}},
  \bibnamefont{and} \bibinfo{author}{\bibfnamefont{S.}~\bibnamefont{Woerner}},
  \bibinfo{journal}{Quantum} \textbf{\bibinfo{volume}{5}}, \bibinfo{pages}{479}
  (\bibinfo{year}{2021}).

\bibitem[{\citenamefont{{Tate} et~al.}(2021)\citenamefont{{Tate}, {Moondra},
  {Gard}, {Mohler}, and {Gupta}}}]{Tate2021}
\bibinfo{author}{\bibfnamefont{R.}~\bibnamefont{{Tate}}},
  \bibinfo{author}{\bibfnamefont{J.}~\bibnamefont{{Moondra}}},
  \bibinfo{author}{\bibfnamefont{B.}~\bibnamefont{{Gard}}},
  \bibinfo{author}{\bibfnamefont{G.}~\bibnamefont{{Mohler}}}, \bibnamefont{and}
  \bibinfo{author}{\bibfnamefont{S.}~\bibnamefont{{Gupta}}},
  \bibinfo{journal}{2112.11354}  (\bibinfo{year}{2021}).

\bibitem[{\citenamefont{Cain et~al.}(2022)\citenamefont{Cain, Farhi, Gutmann,
  Ranard, and Tang}}]{Cain2022QAOA}
\bibinfo{author}{\bibfnamefont{M.}~\bibnamefont{Cain}},
  \bibinfo{author}{\bibfnamefont{E.}~\bibnamefont{Farhi}},
  \bibinfo{author}{\bibfnamefont{S.}~\bibnamefont{Gutmann}},
  \bibinfo{author}{\bibfnamefont{D.}~\bibnamefont{Ranard}}, \bibnamefont{and}
  \bibinfo{author}{\bibfnamefont{E.}~\bibnamefont{Tang}},
  \bibinfo{journal}{arXiv:2207.05089}  (\bibinfo{year}{2022}).

\end{thebibliography}


\end{document}